\documentstyle[prb,aps,multicol,epsf]{revtex}

\begin{document}

\draft
\flushbottom

\title{Electronic Transmission Through Metallic Nanowires:\\
       Generalized Scattering Matrix Approach}

\author{J. A. Torres\footnote{Corresponding author. Email:jtorres@jrcat.or.jp}} 

\address{Joint Research Center for Atom Technology (JRCAT-ATP), Tsukuba, 
Ibaraki 305-0046, Japan.}

\author{J. J. S\'{a}enz}

\address{Dep. de F\'{\i}sica de la Materia Condensada, Univ. Aut\'{o}noma
de Madrid (UAM), E-28049 Madrid, Spain.}

\date{\today}

\maketitle

\begin{abstract}
An easy to implement and powerful method for the solution of 3D scattering 
problems that can be well described by Helmholtz equation is presented. 
The matrix algebra used provides excellent stability versus
the number of junctions as well as great computational speed. 
The matrix truncation method yields an easy single-parameter convergence procedure.
Subsequently, some aspects of the electronic transport through metal nanowires are 
studied by the use of Landauer's scattering approach to the conductance. 
We predict the existence of current vortex-rings patterns due to sharp enough
narrow-wide connections in atomic size point contacts. 
Longitudinal resonances between scattering centers provide a simple physical picture 
for the understanding of negative
differential resistance in ideal monoatomic contacts. Relatively long nanowires with 
high geometrical perfection -like those recently observed by Transmission Electron 
Microscopy- are modelled exhibiting resonant tunnelling and total 
reflection at given incident energy intervals.
\\
\end{abstract}



\begin{multicols}{2}
\narrowtext

Future daily-life technology will be greatly based on miniaturization of both
electronic components and optical devices. State-of-the-art research has already
prototyped tools using simultaneously both electrons and photons confinement. For
instance, in artificially created electromagnetic micro-cavities
interaction light-matter can best be described by new concepts like those of ``cavity
polaritons''\cite{Weisbuch92}, devices in which the light emission process can be
extra-ordinarily fast. Also, quantum well lasers utilizing two-dimensional electron
gas and optical modulators with quantum confinement effect are now practically used.
The ultimate integration of any opto-electronic device has to overcome the
theoretical and practical problem of the waveguiding at tiny sizes for both electrons
and photons\cite{Burstein95}.
Consequently, it is not surprising the huge research effort devoted, in the last
decade, to these waveguides themselves: {\em metallic wires at the nanometer scale}
and {\em optical waveguides at the micrometer scale}. A common characteristic lays on
the basis of these waveguide systems: elastic scattering processes are dominant for
transport. Metallic atomic-size wires and micro-optical guides present lateral
dimensions of the order of the wavelength associated to the particles passing
through. Helmholtz equation governs elastic flux of waves through a general dispersive
system (note here that the independent free-electron stationary Schr\"odinger equation
is equivalent). Therefore, the ability to calculate solutions for this equation is of
great importance. 

As mentioned above, the experimental and theoretical study of metallic nanowires, 
namely their electrical and structural properties, has been focus of intense 
research in the last decade (see e.g articles in Ref. \cite{Serena97}
as well as the various experimental works of Ref. \cite{Experiments}, and the 
theoretical works in \cite{Theory}).
However, only very recently, the combination of Scanning Tunneling Microscope 
(STM) with Transmission Electron Microscopy (TEM) has achieved atomic-resolution
images of metallic 
nanowires\cite{Kizuka97,Kizuka98a,Kondo97,Ohnishi98,Kondo00,Erts00,Rodrigues00}. 
In particular, ultrathin gold wires have been observed exhibiting a striking 
geometrical regularity in some cases.
Here we intend to look at
some possible effects due to the ideal geometry of these nanowires.
Due to the simple -therefore general- phenomenology studied, the same
physics is expected to be found by more sophisticated approaches.

The paper is organized as follows: first we present a Generalized 
Scattering Matrix method for the solution of Helmholtz (or Schr\"odinger) equation
in section \ref{secSM}.
Though the formalism developed can be applied to general scalar wave propagation 
problems and even scattering of light in some cases, 
we use it later to address the problem of conduction through atomic-size contacts:
In section \ref{secWires} we focus on the study of electronic conduction through
metallic nanowires with several models. Finally, in section \ref{secConcl} we 
summarize our results. 

\section{Scattering Matrix Method} \label{secSM}

As a start, let us write the 3D time-independent Schr\"odinger equation 
in $\lambda_F$ and $E_F$ (Fermi wavelength and Fermi energy)
units:

\begin{equation}
 \nabla ^2 \Phi (x,y,z) + 4 \pi^2 \left[ E - V(x,y,z) \right] \Phi(x,y,z) = 0
 \label{eq1}
\end{equation}

A general waveguide geometry can be formed by discretization in 
{\em constant section slices} (where the confining potential $V(x,y,z)$ 
only depends on $x,y$). For a quantum waveguide of {\em constant section},
the solution of the time-independent Schr\"odinger equation is separable
as: $\Phi(x,y,z) = Z(z) \psi(x,y)$, being the separated equations:
\begin{equation}
 Z'' + k^2 Z = 0
 \label{eq2}
\end{equation}
\begin{equation}
 \nabla ^2 \psi + [ (4 \pi ^2 E - k^2) - 4 \pi ^2 V(x,y) ] \psi = 0
 \label{eq4}
\end{equation}
where the longitudinal wavevector $k$ is the arbitrary separation constant.
The general solution for the propagating part of the wave function is:
$Z(z) = I e^{i k z } + R e^{-i k z }$,
where $I$ and $R$ are arbitrary constants to be determined. In $XY$ we have
a generic quantum well determined by the confinement potential $V(x,y)$.
We will assume that this well has some ``hard wall'' boundary conditions at
a given limit. The discrete set of eigenvalues
$ \{ \varepsilon_n \}_{n=0,1,2\ldots}$ (being  $\varepsilon_n \equiv 4 \pi^2 E
- k_n^2 $) can be found analytically or numerically. The known solutions
of equation \ref{eq4} form a complete basis of transversal eigenfunctions
$A \equiv \{ \phi_n(x,y) \} _{n=0,1,2...}$ inside the hard-wall limit, so one
can write the complete solution for the uniform section waveguide as an
expansion in this basis:
\begin{equation}
 \Phi  (x,y,z) = \sum_{n}^{\infty} \phi_n(x,y)
 \left\{ I_n e^{i k_n z } + R_n e^{-i k_n z} \right\}
 \label{eq6}
\end{equation}
where the quantity in between brackets can be seen as the correspondent 
complex coefficients of the expansion. This leads to a vectorial notation in
which the coordinates of $\Phi$ in $A$ are given by $\{ \vec{I}(z) +
\vec{R}(z) \}$, where the $j$-th elements of vectors $\vec{I}$ and
$\vec{R}$ are  $I_j e^{ik_j z}$ and $R_j e^{-ik_jz}$. The basis $A$ is
completely characterized by the confining potential $V(x,y)$.

\subsection{Scattering Matrix of a Single Junction.}

\begin{figure}
\epsfxsize=\columnwidth
\epsffile{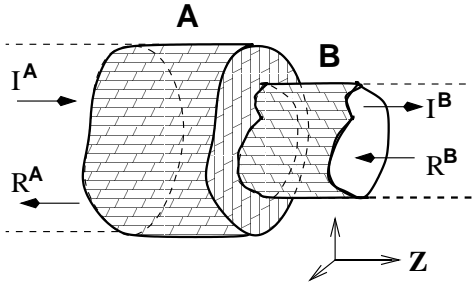}
\caption{Generic single junction discontinuity in a 3D quantum waveguide. Inside
the hard-walls, represented here, the confining potential $V(x,y)$ is arbitrary.}
\label{Fig1}
\end{figure}

Mode-matching techniques 
can be used to obtain the scattering matrix (SM)
for the simplest non uniform system: a single junction. A single junction
discontinuity is shown in Fig. \ref{Fig1}, the general solution and
derivatives in the two uniform waveguide sections are given, in matrix
notation, by:
\begin{equation}
 \begin{array}{lll}
  \Phi ^A(x,y,z) = \vec{I} ^A (z) + \vec{R} ^A (z) & , & \mbox{in basis A} \\
  \Phi ^B(x,y,z) = \vec{I} ^B (z) + \vec{R} ^B (z) & , & \mbox{in basis B}
  \end{array}
 \label{eq7}
\end{equation}
\begin{equation}
 \begin{array}{l}
  \frac{\partial \Phi ^A} {\partial z} = K^A \{ \vec{I} ^A(z) - \vec{R} ^A(z) \} \\
  \frac{\partial \Phi ^B} {\partial z} = K^B \{ \vec{I} ^B(z) - \vec{R} ^B(z) \}
 \end{array}
 \label{eq8}
\end{equation}
where diagonal matrix $K^A$  is given by 
$( k_j^A )^2 = - \varepsilon _j^A + 4 \pi ^2 E $ (analogous definition for $K^B$).

Continuity condition for the wave function and its normal derivative at the junction 
leads to these following two matrix equations:
\begin{equation}
 \vec{I} ^A + \vec{R} ^A = C( B \rightarrow A) \{ \vec{I} ^B + \vec{R} ^B \}
 \label{eq10}
\end{equation}
\begin{equation}
 C^T(B\rightarrow A) K^A \{ \vec{I} ^A - \vec{R} ^A \} 
 = K^B \{ \vec{I} ^B - \vec{R} ^B \}
 \label{eq11}
\end{equation}
Here, matrix $C$ characterizes the intermode scattering or mode coupling due to
the step discontinuity and is defined by overlap integrals. The 
{\em change of basis matrix} to change from basis $B$ to basis $A$, 
$C(B \rightarrow A)$, can only be defined if the eigenvectors of the basis $B$ 
can be expanded in terms of basis $A$; this implies that the definition
domain for $A$ should include the one for $B$ i.e., in geometrical terms, the
hard-wall closed line of $B$ is completely inside the one for $A$ (see Fig. 
\ref{Fig1}).

The element $(j,m)$ of the change of basis matrix $
C(B \rightarrow A)$ is given by:
\begin{equation}
 C_{jm}(B \rightarrow A) = \int_{S_A}dx dy \phi _m^B (x,y) \phi _j^A (x,y)
 \label{eq12}
\end{equation}
where $S_A$ is the area in the plane $XY$ enclosed by the hard-wall limit
correspondent to $A$. From matrix equations \ref{eq10} and \ref{eq11} the 
generalized SM defined as:
\begin{equation}
\left(
 \begin{array}{l}
  \vec{R} ^A  \\ 
  \vec{I} ^B 
 \end{array}
 \right) = \left(
 \begin{array}{cc}
  S_0  & S_1 \\  
  S_2  & S_3  
 \end{array}
 \right)
 \left(
 \begin{array}{c}
  \vec{I} ^A \\
  \vec{R} ^B
 \end{array}
 \right)
 \label{eq13}
\end{equation}
can be readily obtained and is given by:
\begin{equation}
 S_0 = -2 C(B \rightarrow A) M H_{BA} - 1
 \label{eq14a}
\end{equation}
\begin{equation}
 S_1 = C(B \rightarrow A) \{ M N + 1 \}
 \label{eq14b}
\end{equation}
\begin{equation}
 S_2 = -2 M H_{BA}
 \label{eq14c}
\end{equation}
\begin{equation}
 S_3 = M N
 \label{eq14d}
\end{equation}
where:
\begin{equation}
 1 = \mbox{diagonal identity matrix.}
 \label{eq15a}
\end{equation}
\begin{equation}
 H_{BA} = - K_B^{-1} C^T(B \rightarrow A) K_A
 \label{eq15b}
\end{equation}
\begin{equation}
 M = [ 1 - H_{BA} C(B \rightarrow A) ] ^{-1}
 \label{eq15c}
\end{equation}
\begin{equation}
 N = 1 + H_{BA} C (B \rightarrow A) 
 \label{eq15d}
\end{equation}

The analysis of a single junction similar to that shown in Fig. \ref{Fig1}
with the only difference that now $B$ includes $A$, is completely analogous,
equivalent equations and SMs are given in the next. Following
two matrix equations are equivalent to eq. \ref{eq10} and \ref{eq11}:
\begin{equation}
C(A \rightarrow B) \{ \vec{I} ^A + \vec{R}^A \} =  \vec{I}^B + \vec{R}^B
\label{eq16}
\end{equation}
\begin{equation}
K^A \{ \vec{I}^A - \vec{R}^A \} = C^T(A\rightarrow B)  K^B  \{ \vec{I}^B - \vec{R}^B \}
\label{eq17}
\end{equation}
The same definition (eq. \ref{eq12}) is valid for the {\em change of basis 
matrix} to change from $A$ to $B$. From matrix equations \ref{eq16} and
\ref{eq17} the scattering matrix, defined as stated in eq. \ref{eq13}, is
given by:
\begin{equation}
S_0 = M N
\label{eq18a}
\end{equation}
\begin{equation}
S_1 = -2 M H_{AB}
\label{eq18b}
\end{equation}
\begin{equation}
S_2 = C(A \rightarrow B) \{ M N + 1 \}
\label{eq18c}
\end{equation}
\begin{equation}
S_3 = -2 C(A\rightarrow B) M H_{AB} - 1
\label{eq18d}
\end{equation}
where same definitions \ref{eq15a}-\ref{eq15d} are valid if one
swaps $A$ and $B$ indexes.

\subsection{Scattering Matrix of many-junction structures}

The SM of a composite waveguide structure may be obtained by combining
SMs of the corresponding junctions. The procedure to obtain the SM of a 
composite waveguide structure is described in the following.

\begin{figure}
\epsfxsize=0.95\columnwidth
\epsffile{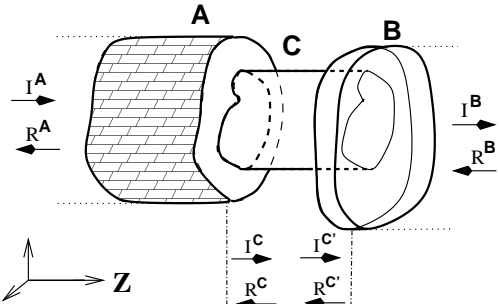}
\caption{Building block general geometry for a composite waveguide structure.}
\label{Fig3}
\end{figure}

Let us assume two consecutive junctions joined by a uniform waveguide 
section as shown in Fig. \ref{Fig3}.
Individual SM for both single junctions have been obtained as described above
and are defined by:
\begin{equation}
\left(
 \begin{array}{l}
  \vec{R} ^A  \\ 
  \vec{I} ^C 
 \end{array}
 \right) = \left(
\begin{array}{cc}
S_0^A  & S_1^A \\  
S_2^A  & S_3^A  
\end{array}
\right)
\left(
\begin{array}{c}
\vec{I} ^A \\
\vec{R} ^C
\end{array}
\right)
\label{eq20}
\end{equation}
\begin{equation}
\left(
 \begin{array}{l}
  \vec{R} ^{C'}  \\ 
  \vec{I} ^B 
 \end{array}
 \right) = \left(
\begin{array}{cc}
S_0^B  & S_1^B \\  
S_2^B  & S_3^B  
\end{array}
\right)
\left(
\begin{array}{c}
\vec{I} ^{C'} \\
\vec{R} ^B
\end{array}
\right)
\label{eq21}
\end{equation}
We will describe now formulae to obtain the total SM of the system
defined by:
\begin{equation}
\left(
 \begin{array}{l}
  \vec{R} ^A  \\ 
  \vec{I} ^B 
 \end{array}
 \right) = \left(
\begin{array}{cc}
S_0  & S_1 \\  
S_2  & S_3  
\end{array}
\right)
\left(
\begin{array}{c}
\vec{I} ^A \\
\vec{R} ^B
\end{array}
\right)
\label{eq22}
\end{equation}
Since the middle connection is a uniform waveguide we can write:
\begin{equation}
\begin{array}{l}
\vec{I}^{C'} = D \vec{I}^C \\
\vec{R}^C = D \vec{R}^{C'}
\end{array}
\label{eq23}
\end{equation}
where $D$ is a diagonal matrix with elements given by:
$ D_j = e^{i k_j^C L} $.
With some matrix algebra from matrix equations \ref{eq20} and \ref{eq21}, 
using \ref{eq23}, one can obtain the wanted SM of the whole composite system:
\begin{equation}
S_0 = S_0^A + S_1^A D S_0^B D M S_2^A
\label{eq25a}
\end{equation}
\begin{equation}
S_1 = S_1^A D [ S_0^B D M S_3^A D + 1] S_1^B
\label{eq25b}
\end{equation}
\begin{equation}
S_2 = S_2^B D M S_2^A
\label{eq25c}
\end{equation}
\begin{equation}
S_3 =  S_2^B D M S_3^A D S_1^B + S_3^B
\label{eq25d}
\end{equation}
where the diagonal identity matrix has been noted as ``1'' and $M$ is:
\begin{equation}
M = [1 - S_3^A D S_0^B D] ^{-1}
\label{eq26}
\end{equation}
This same solution can be expressed in other equivalent ways, nevertheless, 
we found that this is the best formula to reduce computational time since
it requires minimum number of matrix inversions: just one.

For a greater number of junctions system, one starts similarly calculating 
each single-junction SM. Then, the iterative use of this combining recipe 
leads to the total SM.

\subsection{Cylindrically symmetric systems}

In the case of cylindrical symmetry, overlapping integrals get easier since can be
reduced to one dimension. Let us detail the expressions for the simple case of 
plane hard-wall confinement. 

The symmetry assures that will be no matching of modes belonging to different 
acimutal quantum numbers $m$\cite{Morse53}, so we obtain a 2D equivalent problem for
each $m$. The number of 2D problems that one needs to solve increases with tubes
width. 

Let us consider a {\em cylindrical uniform} waveguide section. Electrons are
confined with hard-walls at distance $R$ of axis $Z$, the confining potential
$V(\vec{r})$ is only function of the distance to this axis. Thus, eq. \ref{eq1}
can be separated in cylindrical coordinates, so that the wavefunction can be
written as:
\begin{equation}
\Phi^R(r,\theta , z) = \sum_{m=0}^{\infty} \Phi_m^R(r,\theta,z)
\label{eq30}  
\end{equation}
where:
\begin{equation}
\Phi_m^R(r,\theta,z) = e^{i m \theta}\sum_{l=1}^{\infty} \xi_{ml}^R(r) F_{ml}^R(z)
\label{eq31}
\end{equation}
being $\xi_{ml}^R(r)$  and $F_{ml}^R(z)$ to be specified below.

Due to the cylindrical symmetry, the problem can be considered separately for each
value of $m$; {\em therefore, in the following, we will focus in the problem for a
fixed $m$}.

Functions $\xi_{ml}^R(r)$ are the solutions of the separated radial equation:
\begin{equation}
\left\{
\begin{array}{l}
\frac{1}{r} \frac{d}{dr} \left[ r \frac{d}{dr} \xi_{ml}^R(r) \right] + \\
\, \, \, + \left\{ 4 \pi^2 [E - V(r)]  - [ k_{ml}^2 + \frac{m^2}{r^2}] \right\}
\xi_{ml}^R(r) = 0 \\ \nonumber
\mbox{with the boundary condition: } \xi_{ml}^R(R) = 0
\end{array}
\right.
\label{eqradial}
\end{equation}
Let us consider the basis of these radial eigenfunctions defined between $0$ and $R$:
$B_R(r) = \left\{ \xi_{ml}^R(r)\right\}_{l=1,2,\ldots,\infty}$. 
Then, the wavefunction $\Phi_m^R(r,\theta,z)$ can be expressed as an expansion in
the basis $B_R(r)$ with some coefficients that depend on $\theta$ and $z$, 
written as $e^{i m \theta} F_{ml}^R(z)$ above. In matrix notation this is:
\begin{equation}
\Phi_m^R(r,\theta,z) = \left(
\begin{array}{c}
\vdots\\
{\bf I}_{ml}^R(\theta,z)  \\
\vdots
\end{array}
\right) + \left(
\begin{array}{c}
\vdots\\
{\bf R}_{ml}^R(\theta,z)  \\
\vdots
\end{array}
\right)
\label{eq39}
\end{equation}
where:
\begin{equation}
{\bf I}_{ml}^R(\theta,z) = e^{im\theta} I_{ml}^R e^{ik_{ml}(R)z}
\label{eq40}
\end{equation}
\begin{equation}
{\bf R}_{ml}^R(\theta,z) = e^{im\theta} R_{ml}^R e^{-ik_{ml}(R)z}
\label{eq41}
\end{equation}
being $k_{ml}(R)$ the $z$ component of the wavevector for the subband $ml$.

Therefore  given the (infinite) set of complex numbers  
$ \{ R_{ml}^R, I_{ml}^R \}_{l=1,2,\ldots}$ and given the basis $B_R(r)$, 
we have completely described the wavefunction $\Phi_m^R(r,\theta,z)$. 
For any other radius $R'$ the solution is, obviously, analogous.

This formulation of the problem gives the correspondent SM for each $m$ quantum
number. Matrix equations are the same than in previous sections if we specify the
change of basis matrix  $C$ and the transverse wavevector diagonal matrix $K$. So
that for a single junction joining tubes of radius $A$ and $B$, being $B$ lower than
$A$ (similar to that represented in Fig. \ref{Fig1}) we have:
\begin{equation}
C_{hj}^m(B\rightarrow A) = \int_{0}^{B}  \xi_{mj}^A(r) \xi_{mh}^B(r) r dr  
\label{eq42}
\end{equation}
\begin{equation}
[k_{ml}(R)]^2 = 4 \pi^2 E  - [q_{ml}(R)]^2    
\end{equation}
being $q_{ml}(R)$ given by the transversal quantized levels.

The calculation of the relevant integral, for the {\em simple case of hard-wall 
boundary conditions ($V=V_0$ constant inside the tubes)}, yields:
\begin{equation}
\begin{array}{l}
C_{hj}^m(B\rightarrow A) = \\ \nonumber
 - \frac{\alpha_{mh}}{2} N_{mj}^A N_{mh}^B 
\frac{J_m(\alpha_{mj}B/A) \left[ J_{m-1}
(\alpha_{mh})-J_{m+1}(\alpha_{mh}) \right]}{[ \alpha_{mh}/B]^2 
- [ \alpha_{mj}/A]^2} 
\end{array}
\end{equation}
In this case, radial wavefunctions are given by:
\begin{equation}
\xi_{ml}^R(r) = \frac{\sqrt{2}}{R J_{m+1}(\alpha_{ml})} J_m(r \alpha_{ml}/R)
\label{eq32}
\end{equation}
being $\alpha_{ml}$ the zero number $l$ of $J_m(x)$, the Bessel function of
integer order $m$, and $k_{ml}(R)$ the $z$ component of the wavevector given by:
$ k_{ml}(R) = \sqrt{ 4 \pi^2 [E - V_0] - [\alpha_{ml}/R]^2 }$.

\subsection{Conductance Calculations}

Within this framework, one can easily evaluate the conductance of a two terminal
system on the basis of Landauer's scattering approach \cite{Landauer,Buttiker}, which is 
given by the formula:
\begin{equation}
G = G_0 \sum_{n} T_n
\end{equation}
where the sum runs to all propagant modes in the input lead (propagant modes are
defined as those with a correspondent real $k$, while the rest -those with
imaginary longitudinal wavevector $k$- are called evanescent). 
Here, $G_0$ is the universal spinless quantum of conductance with the value $G_0 =
2e^2/h$. The transmission probability for incidence through mode $n$
can be easily obtained and is given by:
\begin{equation}
T_n = \sum_j \left| \left(  S_2 \right)_{jn}  \right| ^2  \frac{k_j^f}{k_n^i}
\end{equation}
where $\left(  S_2 \right)_{jn} $ is the $(j,n)$ element of the $S_2$ matrix
given in equation \ref{eq25c}, the sum runs to all propagant modes of the final 
lead, $k_j^f$ and $k_n^i$ are the $j$ and $n$ longitudinal wavevectors of the 
final and initial tubes respectively.

\subsection{The truncation method}

One crucial point in the correct calculation of the SM of a quantum waveguide 
junction is the right truncation of two infinite series of modes representing 
wave functions at both sides of the junction. 

In the analysis of junctions, different converged results may be obtained for
different ratios of the number of modes retained on each side. This convergence
problem is usually referred  to as relative convergence phenomenon
\cite{Shih89,Mittra71}.  Theoretical studies have found  that the correct converged
result is obtained if the ratio of modes on each side is taken the same as the
ratio of  the corresponding waveguide  widths \cite{Mittra71}. Further
calculations give another optimum value for this ratio which is 1.5 times the
ratio of waveguide widths \cite{Weisshaar91}. 

In our analysis we have found that one can generalize this optimal ratio
criterium to  an optimal ``procedure'' which gives different ratios in different
cases. Our approach is based on observations of the coupling between different
modes at waveguide junctions. 

\begin{figure}
\epsfxsize=\columnwidth
\epsffile{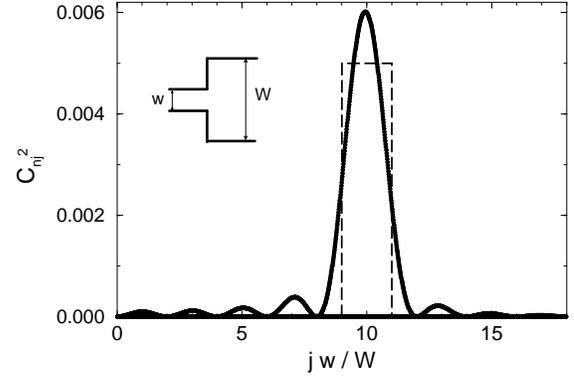}
\caption{Square modulus of the overlapping integral $C_{nj}(w \rightarrow W)$
versus the wide mode index $j$. The values of the widths are $w/\lambda_F =1.2$,
$W/\lambda_F =200.2$ and the fixed value $n$ taken is $n= 10$ (see text for more details).}
\label{Fig4}
\end{figure}

For simplicity, let us consider the 2D axial symmetric  hard-wall problem of a
single junction like the one shown in the inset of Fig. \ref{Fig4}. Assuming
incidence from left to right and for a fixed incident mode  $n$, the plot of the
coupling integral $C_{nj}(w \rightarrow W)^2$ versus  ``wide'' transversal
modes, $j$, is ``peaked'' at $j=v$ such that its  transverse wave vector
satisfies $q_v \approx q_n$. Thick dots in Fig. \ref{Fig4} represent values of
this overlaping integral, the peak can be clearly seen at $q_v \approx q_n$ (in
this simple case  $q_n= n \pi / w$ and $q_v= v \pi / W$). In a ``mean field
approximation''  \cite{Szafer89} true overlaps are approximated by a uniform
coupling to all modes
within one level spacing, represented by the dashed line in Fig. \ref{Fig4}.
This approximation gives excellent total transmission results when the difference
in widths is big \cite{Saenz01}.  
We can conclude that exists a given band of modes of the wide side effectively 
coupled to the $n$-th mode of the narrow side. One needs to take into account all 
modes in this continuous interval to assure a correct representation of the 
``narrow'' mode $n$ in the ``wide'' basis of transversal eigenfunctions. The 
truncation recipe that we have found is based on this observation. 

Though we have shown here a simple case, this mode matching scheme holds for
different confining potentials and presumably for all of them (in particular we
have checked some axial symmetric potentials in 2D and 3D: hard-wall, soft wall
and harmonic. Also some non symmetric 2D and 3D potentials have been checked:  
hard-wall and soft-wall)
Consequently, for one single junction similar to that in Fig. \ref{Fig1}, the 
truncation procedure is as follows. Given transversal levels sorted in crescent 
energy: 
(i)
Take one arbitrary number of modes $N$ in the narrow side, greater or 
equal than the number of propagant ones. This implies that $N$ can be
written as $N = N_p + N_e$, where $N_p$ is the number of propagant 
modes and $N_e$ is the number of evanescent modes in the narrow tube. 
(ii)
Take all modes of the wide side which are effectively connected 
to the ones chosen in the narrow side. This implies that we should use 
every wide-mode whose energy  lies bellow $E_{max}$ given by 
$E_{max}=E_{N+1}$. In other words, $E_{max}$ is the energy correspondent 
to the $N+1$ level of the narrow side.

For the truncation in the case of a greater system, in which the number of 
junctions is larger, the procedure is the same but choosing the narrowest of
the whole  system (i.e., the one that has the sparser transversal energy levels) 
as the tube to define the threshold transversal energy $E_{max}$. This $E_{max}$ 
value is then used to fix the number of modes retained at the rest of the 
uniform waveguide sections of the system.

\begin{figure}
\setlength{\unitlength}{\columnwidth}
\begin{picture}(1,1.21)(0,0)
 \put(0.46,0.88){ \epsfxsize=0.4\columnwidth
                 \epsffile{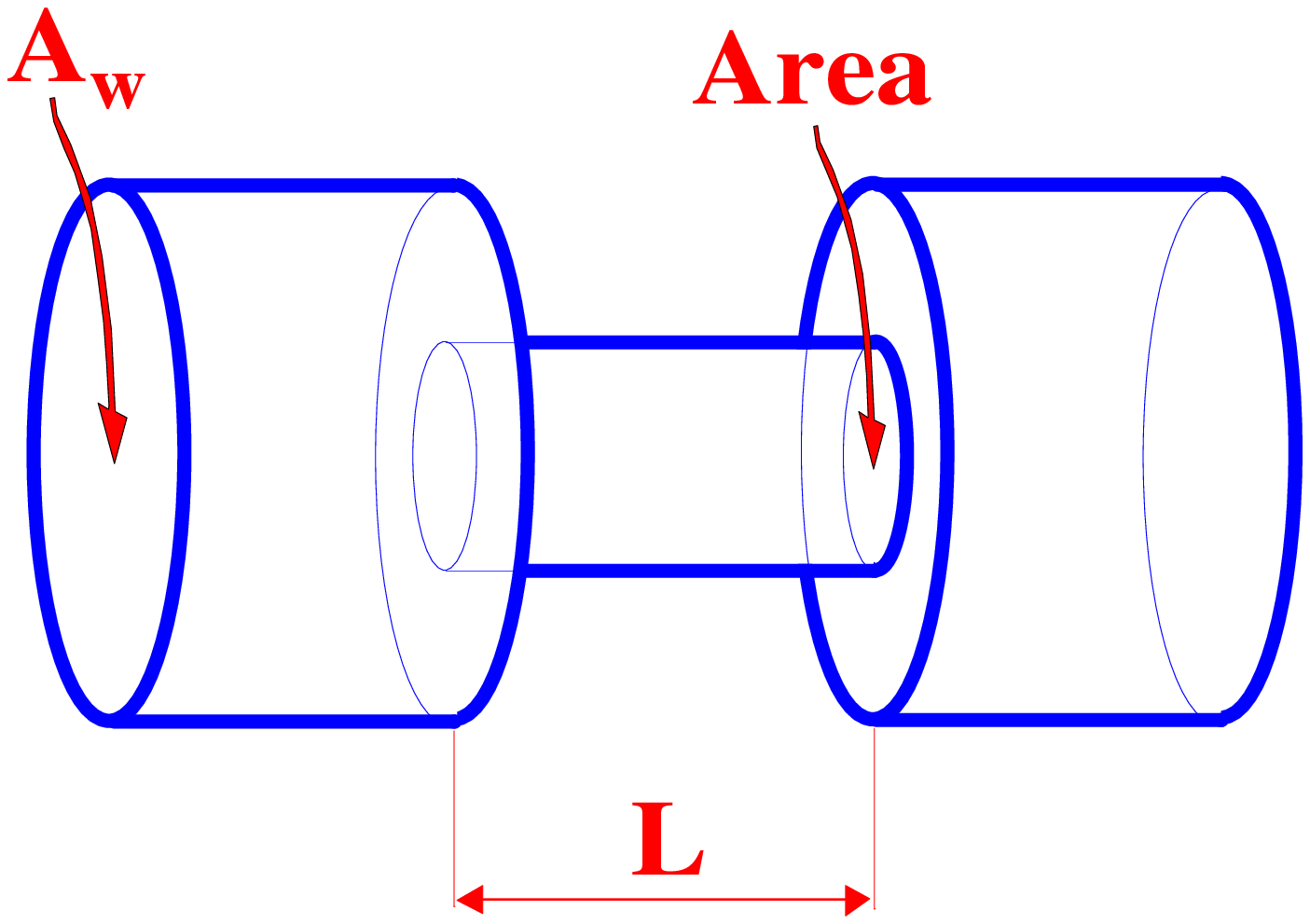} }
 \put(-.02,0){\epsfxsize=0.95\columnwidth
             \epsffile{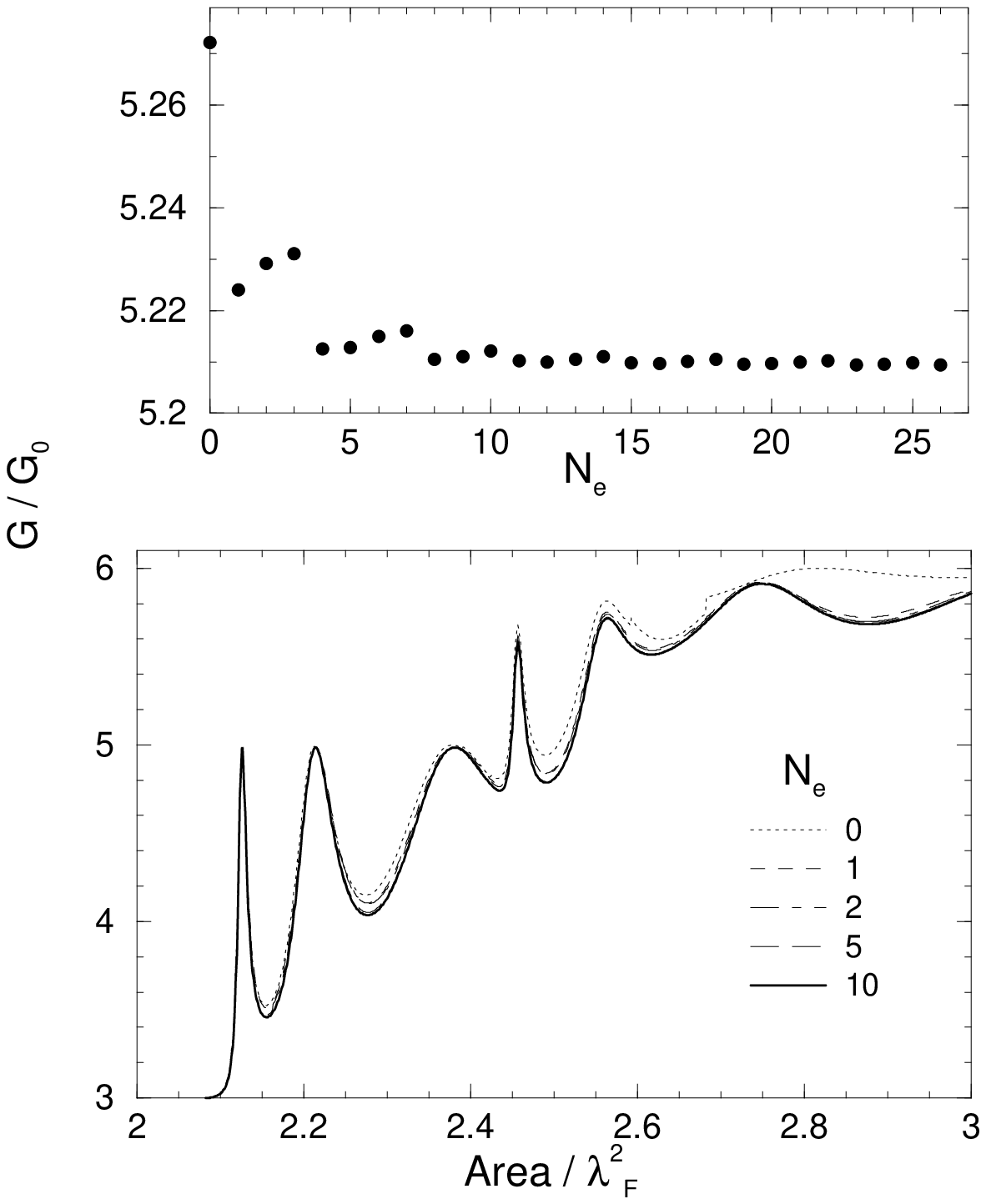}  }
\end{picture}
\caption{Example of convergence tests.
{\bf Bottom}: conductance of a 3D cylindrically
symmetric wide-narrow-wide contact versus the narrow area. Different plots can be
seen for increasing number of evanescent modes retained in the narrow tube, $N_e$.
{\bf Top}: $G$ vs $N_e$ for the worst converging point in the same system:
$A/\lambda_F^2 \approx 2.49$. Parameters used in both plots are: L= 4
$\lambda_F$, $A_w$ = 4 $\lambda_F^2$.}
\label{Fig5}
\end{figure}

Using this procedure, the number of evanescent modes taken in the narrowest tube
$N_e$ is the only free parameter, it can be tuned for proper convergence of the 
results. In Fig. \ref{Fig5} we can see the convergence 
of the conductance versus $N_e$  (note that the number of evanescent modes taken 
into account in the wide side could be quite larger than $N_e$). The plot has been 
done for the 3D ``hard-wall'' problem and is quite representative of the general 
behaviour for different confining potentials.

The problem converges for $N_e$ equal four with an error of $\pm 1 \%$.
The crucial point is that this way of making the truncation ensures  good coupling 
of every mode at the narrow side, and the appropriate description  of these modes 
is the main influence in conductance values.
It is worth to notice here that matrices involved are rectangular in general. 
Algebraical operations done assure the appropriate use of 
them: no numerical inversion of non-squared  matrices has been done. 
One good property 
of this particular way of treating 
matrices involved is that flux is conserved with any truncation 
chosen; this implies that $T_n + R_n =1$ even for $N_e=0$ 
(where $T_n$ is the total transmission and $R_n$ is the total reflection for 
one particular propagant mode $n$ in the initial lead).

By the use of the matrix algebra proposed here, numerical unitarity and stability
are achieved with ease. The SM formalism used, in which information from both ends of
the system is taken in each step, avoids numerical roundoff errors accumulation
in particular when combined with a suitable change of transversal basis as a function
of the local width of the system. The truncation procedure explained above is the key
for the achievement of these characteristics: it produces a self filtering of numerically
unstable components without further data manipulation, hence the unitarity (to compare with
recursion-transfer-matrix methods see for instance Ref. \cite{Hirose95}).

\section{Metallic Nanowires} \label{secWires}

For several years now and in various experimental ways, the formation of ultrathin
wires between two macroscopic pieces of metal has been routinely achieved and characterized 
\cite{Serena97}.  However, direct information about
the atomic structure of these wires has not been available until recently when various experiments have achieved atomic resolution TEM images on gold nanowires 
\cite{Kizuka97,Kizuka98a,Kondo97,Ohnishi98,Kondo00,Erts00,Rodrigues00}.
One of the most interesting features seen
 is the almost ideal perfect geometry, as well as surprising 
helical structures\cite{Tosatti00,Kondo00}. To the date, theoretical work has been 
focused on the  study of the stability and electronic structure for the thinnest 
monoatomic chains \cite{Torres99}; while for the helical wires works have been
focused esentially in the prediction of the very structure for various metals 
\cite{Gulseren98}, and the influence of helicity in the ideal ballistic conductance 
\cite{Okamoto00}. 
In the second part of this work -by making use of the calculation method 
we have just described above- we approach this problem.
We will point out some remarkable effects in the current and conductance patterns 
due to the ideal {\it sharp} geometries observed, these effects are valid in the presence or
not of helicity.   
For a study on the effects of global {\it smooth} shapes in the conductance of atomic-scale
3D contacts see Ref. \cite{Torres94}.

\subsection{Vortex rings in narrow-wide connections.}

\begin{figure}
\setlength{\unitlength}{\columnwidth}
\begin{picture}(1,0.67)(0,0)
 \put(0.44,0.35){ \epsfxsize=0.45\columnwidth
                 \epsffile{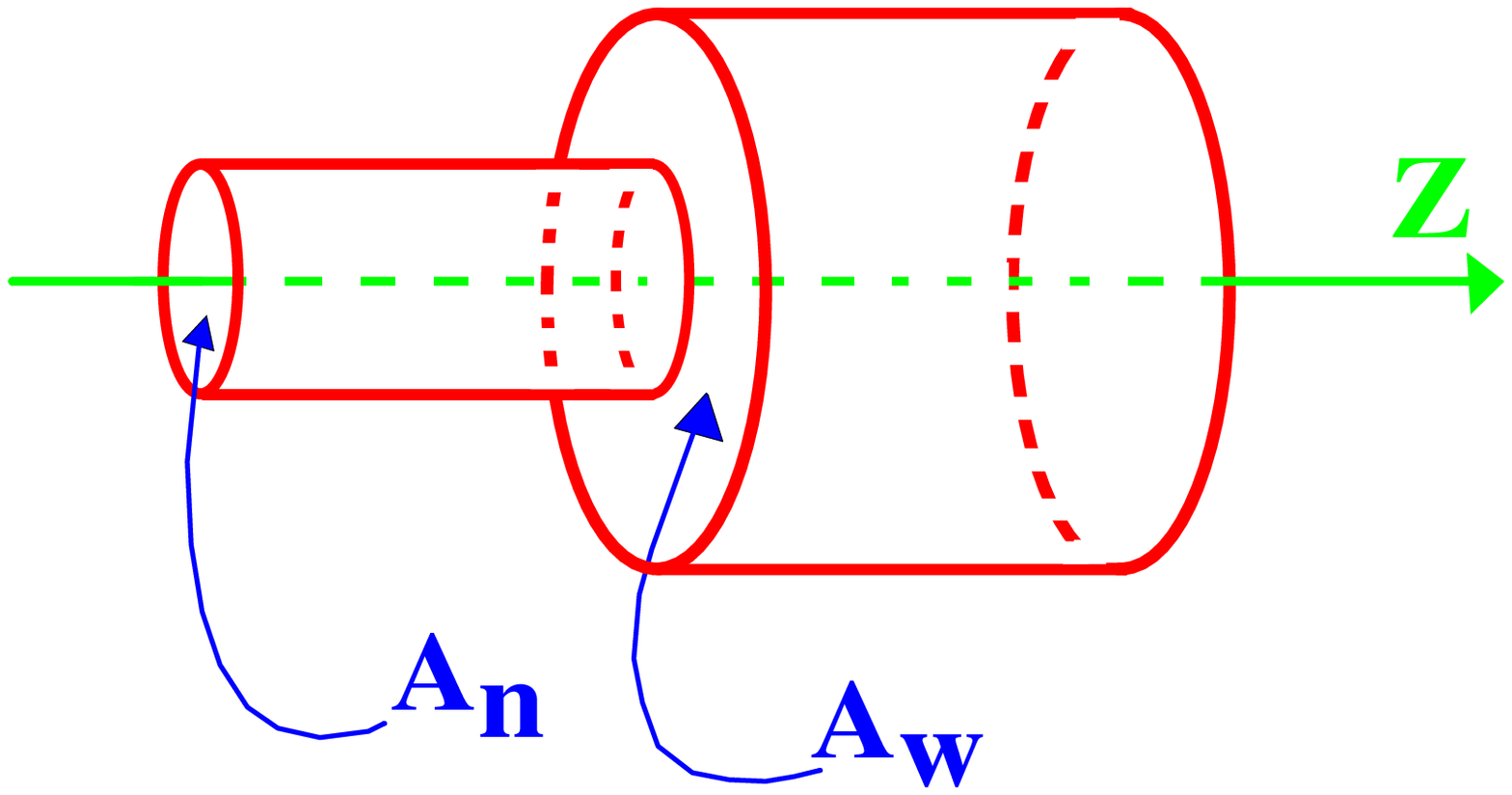}       }
 \put(-.05,0){   \epsfxsize=\columnwidth
                 \epsffile{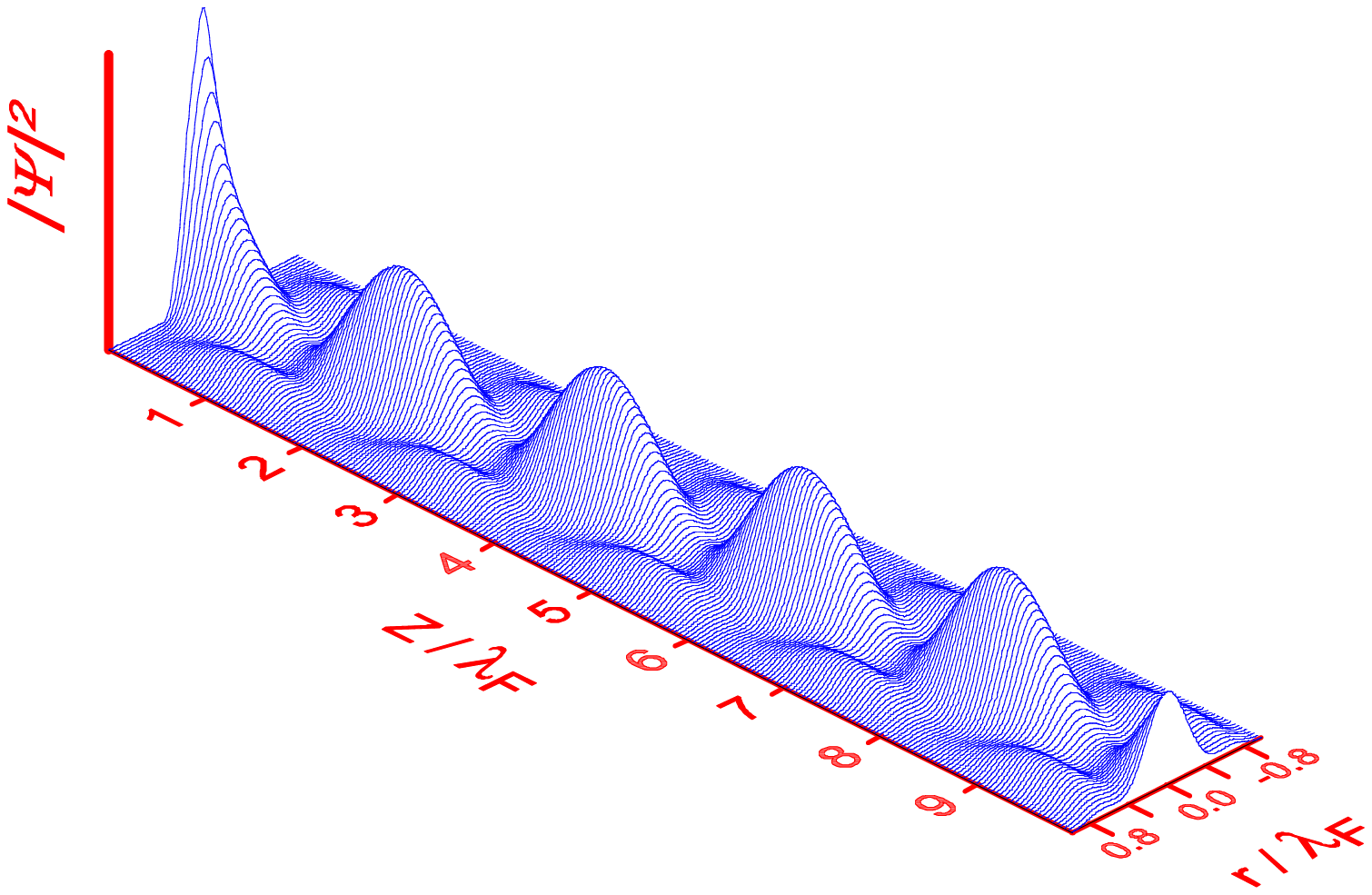}        }
\end{picture}
\caption{Charge density for a NW structure in a longitudinal
cut. The incident narrow tube holds a single propagant mode while two
modes propagate in the wider one. The junction is done at $z=0$ so that,
in the figure,  we can see the charge density in the wider tube.
Parameters used are $An = 0.5 / \lambda_F ^2$, $Aw = 3 /  \lambda_F ^2$.}
\label{fig8b}
\end{figure}

A narrow-wide connection is eventually formed when an unsupported gold 
nanowire is done by experimental means\cite{Kizuka98a,Kondo97,Kondo00}. 
This can be modelled by a 
simple single junction connection in a 3D wire similar to that sketched in Fig. 
\ref{fig8b}.  Previous theoretical work on 2D nanowires have pointed out the
existence of a vortex structure, for the quantum current density, as a result
of a narrow-wide connection. Interestingly, a change of pattern
from laminar to ``turbulent'' flux is obtained for ballistic 2D free-electron
calculations \cite{Berggren92}. In the analogous simplest case for 3D wires we
observe a similar behaviour and transition, the onset of vortex structure
formation is intimately linked to the income of the second transversal level in
the wider part. 

\begin{figure}
\epsfxsize=\columnwidth
\epsffile{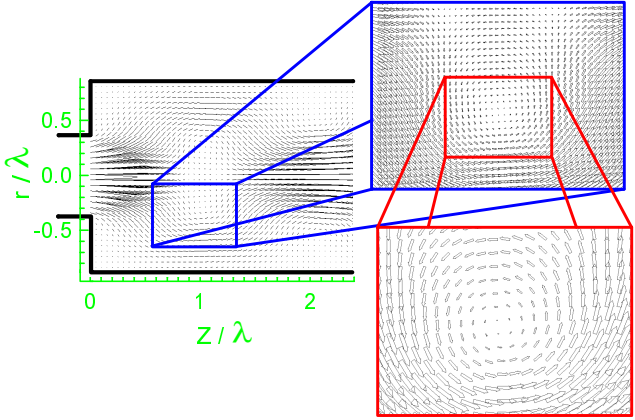}
\caption{Quantum current density map for the same NW structure in
figure \ref{fig8b}. The quantum current opens up as arrives to the wide lead,
further reflection with the walls makes it converge to the centre. This is
periodically repeated ad infinitum. Two sucessive magnifications give
detail of one of the vortex ring formations.}
\label{fig9}
\end{figure}

In Fig. \ref{fig8b} and \ref{fig9} we can see the charge and the current density
for this 3D narrow-wide structure. The width of the narrow wire allows just a
single propagating channel while the wide part allows two channels, this is the
minimum required to exhibit the vortex structure that is remarked in Fig. \ref{fig9}. 
Periodic charge and current density structures are formed along the
final wider tube, this can be readily explained in terms of constructive and 
destructive interference patterns for the only two channels occupied in the wide
lead. There are points with charge density zero regularly distributed at both
sides of the charge density maximums in Fig. \ref{fig8b}, these points are
vortex centers in the quantum current map of Fig. \ref{fig9}. 
Note that, due to the axial symmetry, these vortex points define rings in real
space, and the flux sourrounding them torus surfaces.

\begin{figure}
\epsfxsize=\columnwidth
\epsfysize=0.9\columnwidth
\epsffile{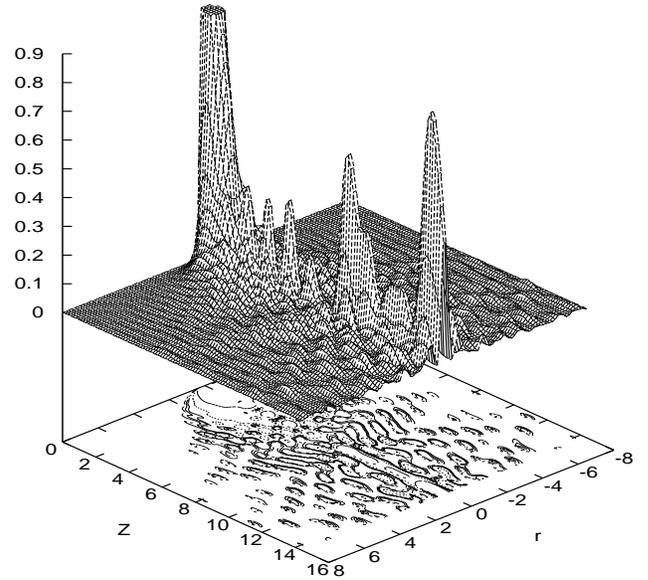}
\caption{Charge density map for a Narrow-MuchWider structure than that
in figure \ref{fig8b}. A longitudinal cut can be seen analogous to that
in figure \ref{fig8b}. Non periodic maxima at the axis can be seen this
time. The disordered pattern is remarked by contour plots at
$| \psi |^2$ values (in axis units): 0.015, 0.02 ,0.025, 0.05 and 0.15.
Parameters used are $An = 0.5 /\lambda_F ^2$, $Aw = 200 /  \lambda_F ^2$.}
\label{fig9b}
\end{figure}

When the wide lead gets much wider, many transversal levels come up and the
charge density structure gets irregular as can be seen in Fig. \ref{fig9b}. 
The same explanation remains valid this time: the pattern results from the 
interference of many -in this case-  longitudinal wavelenghts, 
the period of these patterns increases with the number of
levels involved, eventually it gets seemingly non periodic as the case seen in
Fig. \ref{fig9b} where no regular pattern is observed.

These results could be important for the understanding -and therefore control- 
of coherent waves by tuning geometrical parameters. In particular
the transition from a regular one dimensional lattice of maximums to a disordered pattern
by changing the width of the outgoing waveguide could be amenable of experimental
verification (may be with electromagnetic waves rather than electrons).

\subsection{Resonances inside Wide-Narrow-Wide structures}

A single junction, as the one seen in previous section, can be regarded as a scattering
center in general terms. Realistic wires have atomic structure and therefore many 
possible scattering centers for the conduction electrons passing through. The inclusion 
of increasing number of scattering centers gives rise -as we will see- to a richer phenomenology. 
In this subsection we will focus in the case of just two scattering centers: a wide-narrow-wide
(WNW) junction. Two (strong) scattering centers suffice to predict negative differential 
resistance in point contact experiments.

\begin{figure}
\epsfxsize=\columnwidth
\epsffile{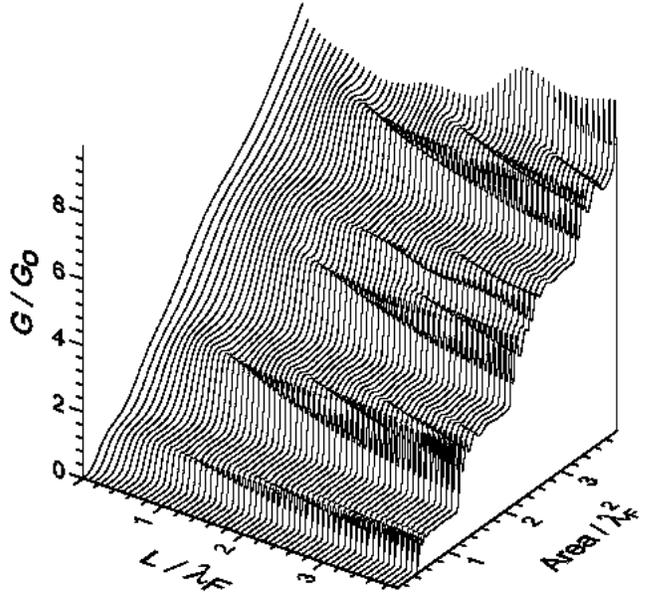}
\caption{Conductance for a 3D WNW cylindrically symmetric contact
(after Ref. [28]).}
\label{fig10}
\end{figure}

The ballistic conductance of a WNW structure, as the one in the inset of fig. \ref{Fig5},
presents a dense resonant behaviour \cite{Torres95} as a function of the narrow area and lenght  
as can be seen 
in fig. \ref{fig10}. 
Being this a very simplistic model, it is a fair representation of what can be expected 
when we have these main ingredients: $i)$ a couple of strong scattering centers at a 
given distance, $L$, and $ii)$ transversal confinement (characterized by the area $A$).
In these cases we can expect a ``wavy landscape'' for the conductance as a function of 
($L$, $A$), though may be not the same than the one in Fig. \ref{fig10}. Clearly, however,
the variation of any of these relevant parameters can eventually give any derivative.
This diversity of results is what -we will see- is possible for the case of monoatomic contacts.

\begin{figure}
\epsfxsize=\columnwidth
\epsffile{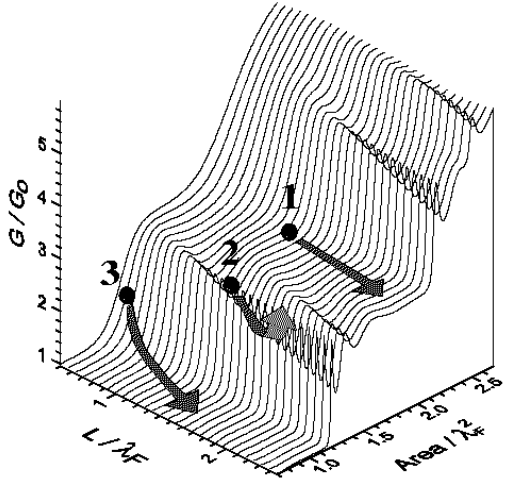}
\caption{Conductance for a 3D WNW contact (magnification). Possible
evolutions for the conductance in nanocontact experiments are indicated with
arrows.}
\label{fig11}
\end{figure}

Within this very simple WNW representation of atomic size contacts, we can easily
locate the zone in Fig. \ref{fig10} that reasonably belongs to monoatomic (or very few 
atoms) contacts for typical metals. This zone is what we have represented 
in Fig. \ref{fig11}, on it we have labeled three different  points  belonging
to 3 different possible particular geometries or ideal contacts.
Arrows indicate the evolution due to an eventual stretching process of these
few-atoms contacts. Before breaking, seems reasonable to think that the
effective lenght will increase and effective area will slightly decrease
\cite{Torres95,Torres96}, this is
what the arrows indicate. Now, as can be seen, as a function of the starting
spot, the subsequent evolution of the conductance can be very different.
If we are initially located in point 1, the evolution will be mostly quantized
(values very near $G_0$) and essentially horizontal; the probability of this
event to happen is quite
high due to the great number of points in the plane ($L$, $A$) belonging to this
type. However, there is a good chance of falling in a place like the one labeled 3, in
this case -as the arrow indicates- we expect to observe a decreasing of $G$ before
breaking the contact. Finally, there is a little but non negligible chance for our
starting point to be like the labeled 2; in this case, since we are located
in the minimum of the ``valley'', further stretching will make an increase in the
conductance as indicated. 

The behaviour for case 2 is the more surprising since it predicts increasing
of conductance while stretching the wire. Should this model be valid, it
could be expected to be found for given geometries and/or metals. In fact there is
compelling experimental evidence of negative differential resistance for the case of
Aluminum\cite{Muller92,Scheer97}, and theoretical explanation in terms of 
atomistic density functional calculations \cite{Sanchez97} for the case of relatively wide
cross-section wires.
Our present  free-electron result provides a helpful simple view of good part of the
relevant physics in these systems, a complementary piece of the whole picture,
especially relevant for the few-atoms-contact case.

\subsection{Resonant Tunneling in periodic WNW structures}

Long physical wires have their thinnest expression in what we can call monoatomic
(in width) long wires. The existence of -relatively long- monoatomic gold wires
has been experimentally reported \cite{Ohnishi98,Rodrigues00,Yanson98} and theoretically studied
\cite{Torres99,Tosatti00,Todorov00} very recently. 
In a chain like this, each atom is a system on its own that holds localized electronic states in
addition to those extended states responsible for conduction.
Very simple modelling of this can be done by
the repitition of a WNW structure. As we will see, this periodic repetition  gives
rise to a long chain of 3D ``resonance boxes''. With the selection of proper
dimensions for the geometrical parameters of these boxes, localized states can be
contained inside. The correspondent energy levels of this states can be an open
door for {\em resonant tunneling} phenomena in these 3D structures.
This is proved here for a sufficiently long periodic wire.

\begin{figure}
\setlength{\unitlength}{\columnwidth}
\begin{picture}(1,0.76)(0.063,0)
  \put(0,0){
    \epsfxsize=\columnwidth
    \epsffile{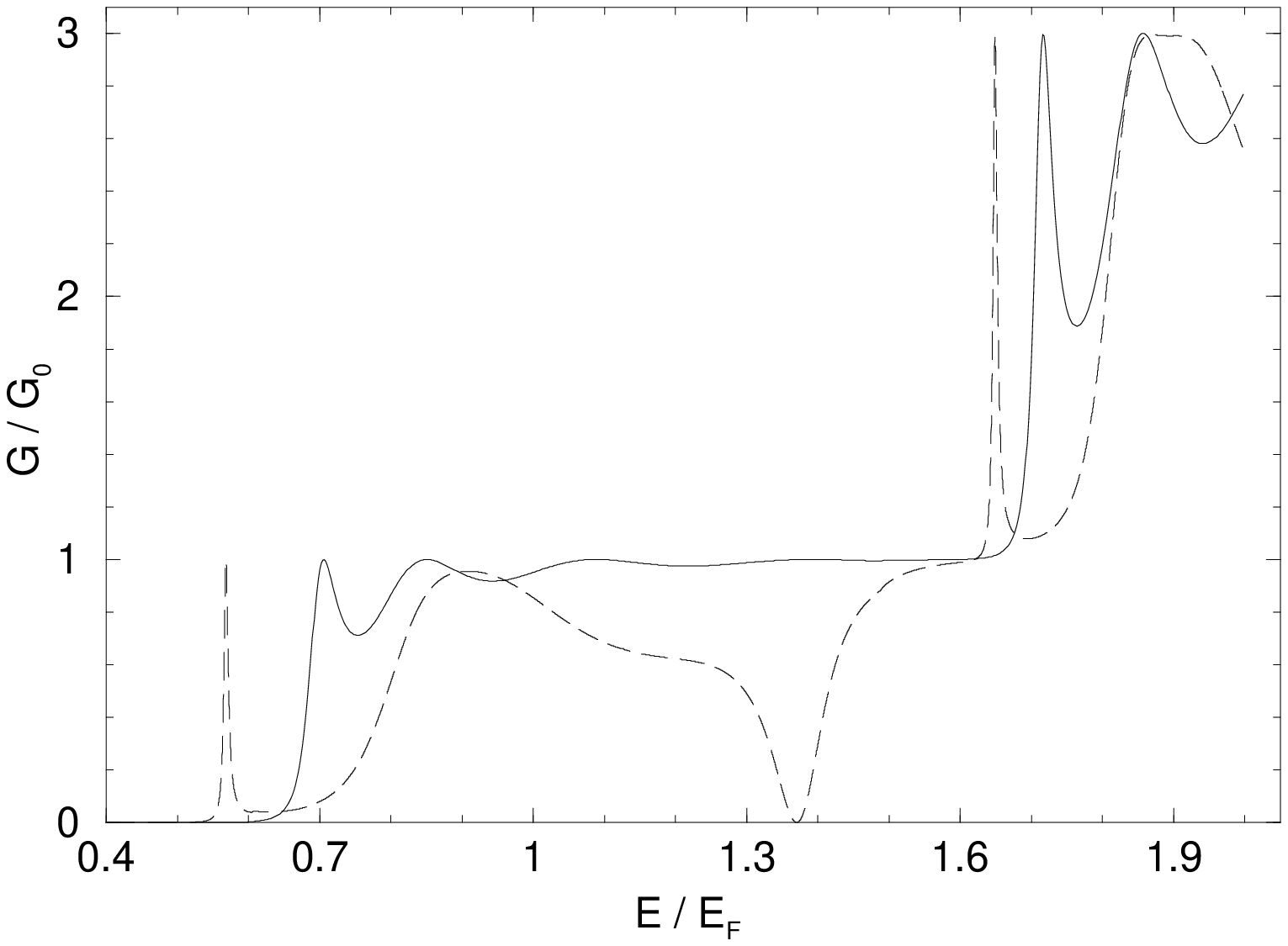}     }
  \put(0.17,0.37){
    \epsfxsize=0.46\columnwidth
    \epsffile{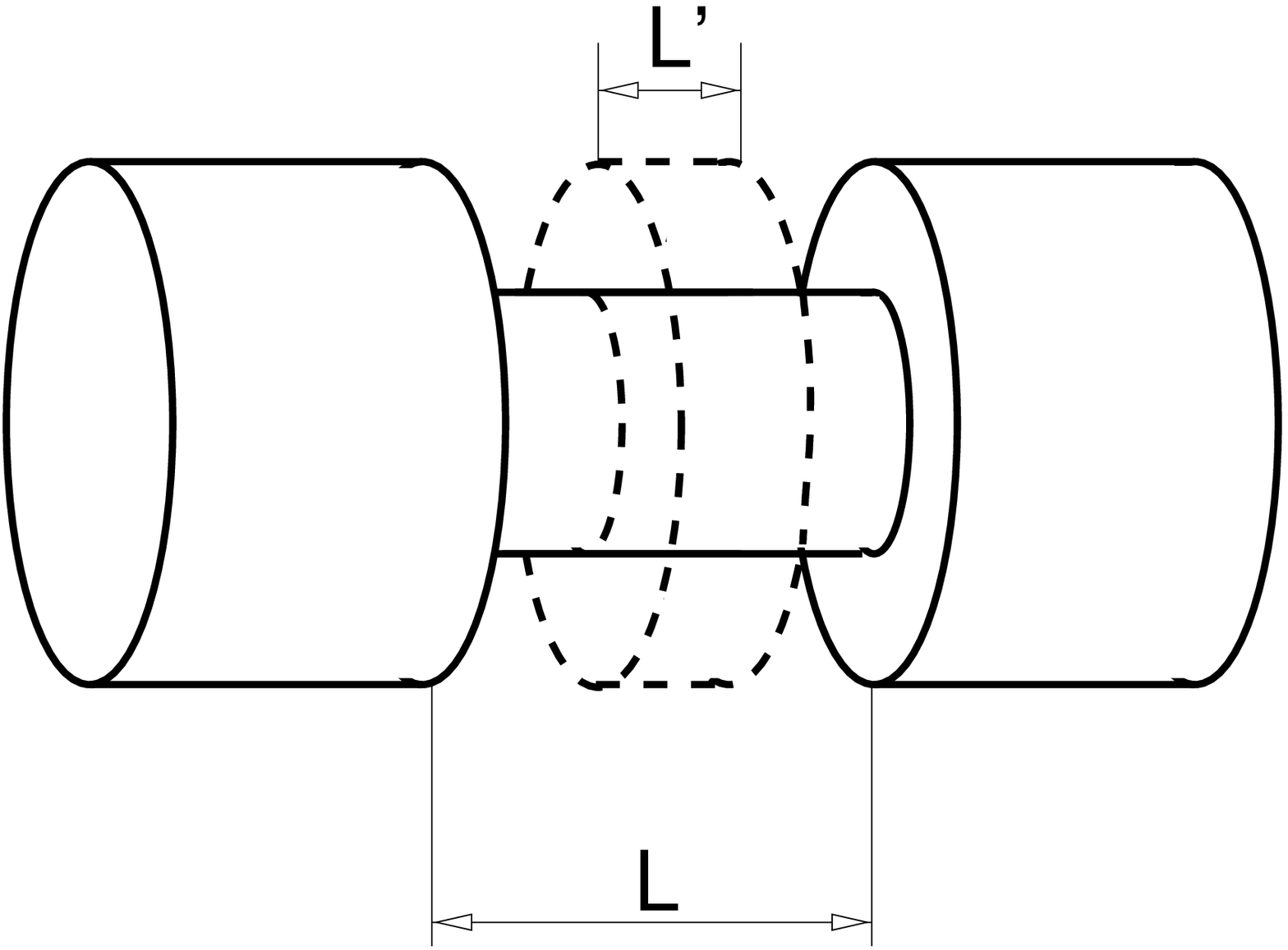}     }
\end{picture}
\caption{Conductance vs incident Energy for WNW (dark line) and a
{\it Resonant Tunneling} single box structure. Both geometries are shown
in the inset. A clear total transmission peak appears before the onset of the
first propagating level inthe narrow tube. As well, the resonant box induces
total inhibition of transmission in the first plateau.}
\label{fig12}
\end{figure}

For a clear exhibition of resonant tunneling phenomena in 3D structures let us
start with the standard WNW problem as shown in Fig. \ref{fig12}. The evaluation
of $G$ versus incident energy yields the expected quantized steps with resonance
oscilations overimposed. There is a threshold energy at $E/E_F \approx 0.7$ marking 
the onset of the first propagating level in the narrow tube, which
is seen as the first jump in conductance. Now, if we mantain the same geometry
but we add -in the central narrow tube- a wider box as shown in the inset, we can
clearly appreciate the appearance of a resonant tunneling peak quite before this 
threshold energy. This is completely analogous to the obtained in 1D and 2D 
models.

\begin{figure}
\epsfxsize=\columnwidth
\epsffile{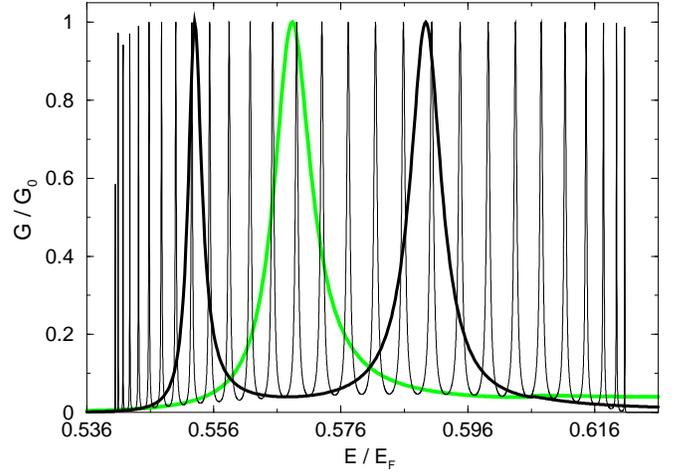}
\caption{Resonant Tunneling peak magnified for 1,2 and 28 boxes (see text).}
\label{fig13}
\end{figure}

It can be clearly prooved that perfect transparence is achieved at the resonant
tunneling energy. Not only the value of transmission gets to 1 at a
given E, but the periodic repetition of this structure does not filter down
conductivity but simply increases the number of peaks. There is a new peak for each
3D quantum resonant box we add: a splitting effect between degenerate localized
levels is proved as can be clearly seen in Fig. \ref{fig13}. In this figure, a zoom
of the resonant peak is seen for the cases of 1, 2 and 28 resonant boxes.

\begin{figure}
\setlength{\unitlength}{\columnwidth}
\begin{picture}(1,0.76)(.065,0)
  \put(0,0){
    \epsfxsize=\columnwidth
    \epsffile{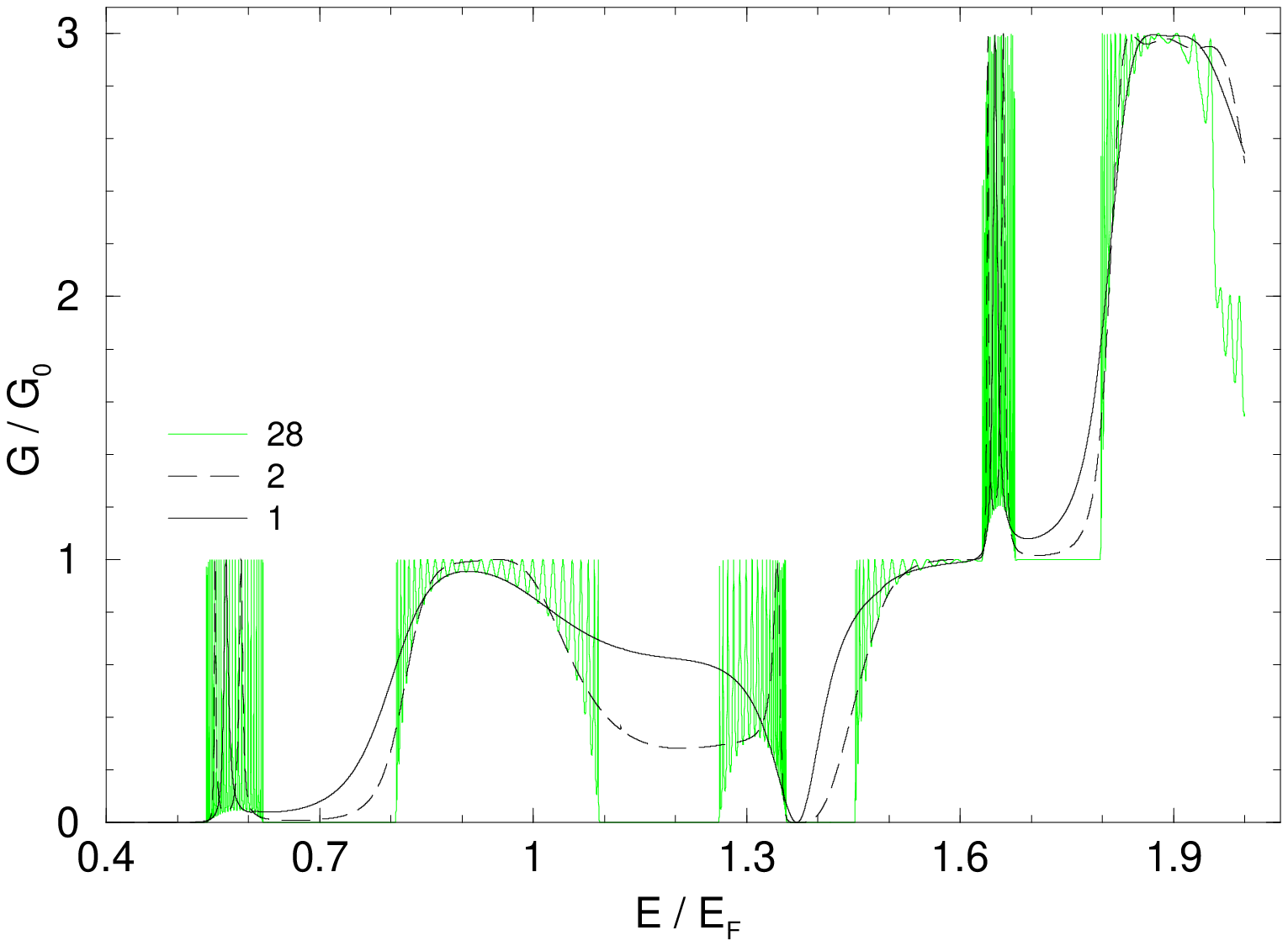}    }
  \put(0.105,0.445){
    \epsfxsize=0.65\columnwidth
    \epsffile{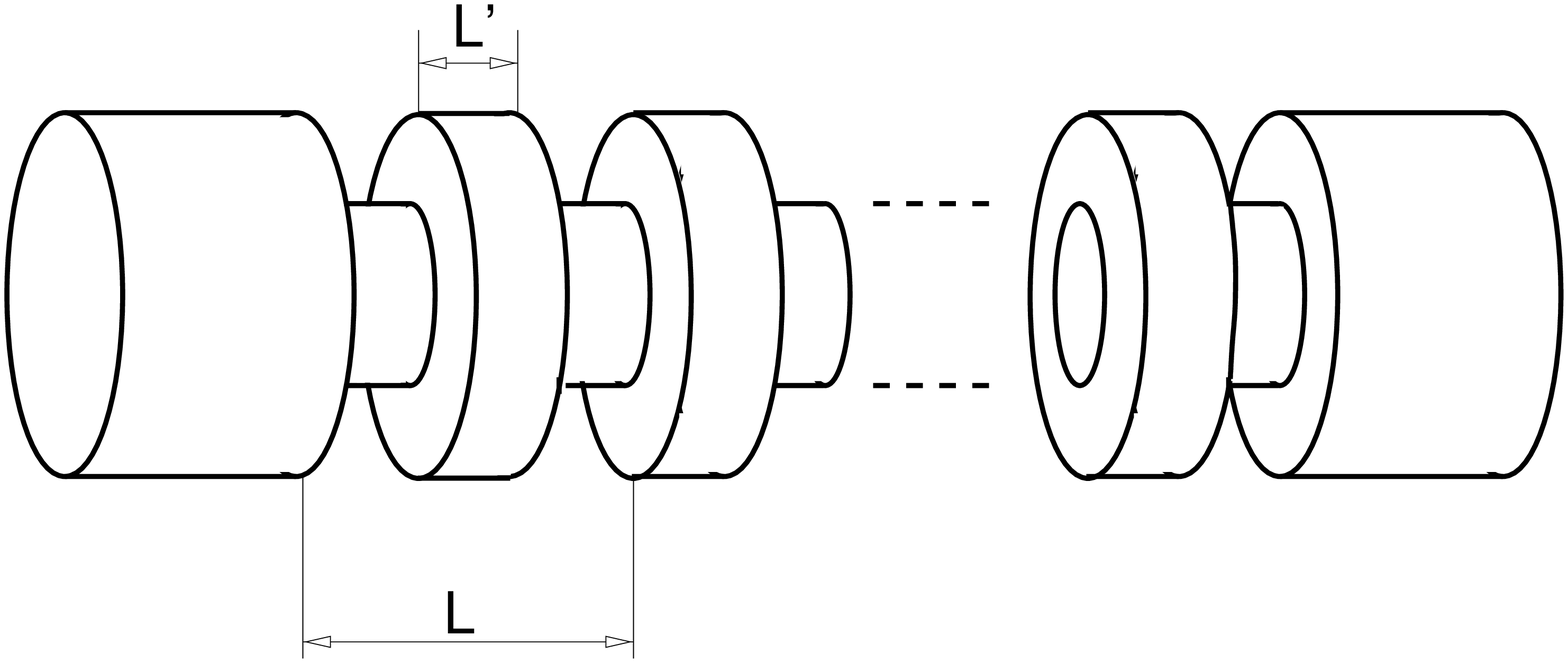}     }
\end{picture}
\caption{Resonant Tunneling, and ``bands'' for 1, 2 and 28 resonance boxes.}
\label{fig14}
\end{figure}

Perfect transmission peaks are not the only  consequence of interference, 
total reflection is also obtained as can be seen in Fig. \ref{fig14}. A ``band like''
structure appears as a consequence of the periodicity imposed in the chain of 28
boxes. There are intervals of energy where transmission is nearly one and
intervals where it is practically zero, this is in connection with conduction bands and
energy gaps of simple one dimensional band theories \cite{Kittel71}. The complete
inhibition of transmission is not obtained in pure 1D problems while for 2D and 3D
problems total reflection is naturally achieved\cite{Sols89}.

\section{Conclusions} \label{secConcl}
We have developed a formulation of the Scattering Matrix method to solve
general three dimensional waveguide problems. The truncation method, based
on the way the mode coupling is done, provides quick convergence of results
with one single parameter to be tuned: the number of evanescent modes in
the narrowmost part of the system. The method has been already successfully
used for the calculation in optical waveguides \cite{Garcia97} and
electronic ballistic contacts\cite{Torres95,Torres96,Pascual97,Garcia96}.
Examples of convergence tests and calculations have been
also shown. In particular, we have studied with some detail the problem
of quantum conductance in metallic nanowires. Various simple models of
crescent number of junctions have been studied. For narrow-wide connections
charge density and quantum current patterns have been shown, interestingly
vortex rings are obtained for this simple geometry in analogy to previous
2D calculations. This vortex structure starts forming a regular pattern
and becomes apparently disordered with increasing tubes width. The calculation for
wide-narrow-wide connections gives rise to strong resonant behaviour of
the conductance as a function of width and length of the contact. Due to
the very recent appearance of experiments in which impressively straight
nanowires have been imaged, seems pertinent a closer look at this idealized
geometry model. Therefore, interpretation of WNW results, leads to the
prediction of several behaviours of the conductance in the last plateaus of
an experimental breaking proccess; in particular negative differential
resistance can be expected. Further addition of junctions to form periodic
WNW structures gives rise to Resonant Tunneling effects in what we can think
of as a simple model for a long periodic nanowire. Opossite to 1D models and
analogous to 2D, not only perfect transmission is obtained for the resonant
tunneling energy but perfect reflection appears for some finite intervals
of energy.

\section{Acknowledgements}
The authors are grateful to M. Nieto-Vesperinas, A. Garc\'{\i}a-Mart\'{\i}n, 
E. Tosatti, J. J. Kohanoff, M. Okamoto, T. Uda, K. Takayanagi and K. Terakura 
for enlightening discussions. 
One of the authors (JAT) acknowledges funding support from: European Union through 
TMR grant ERBFMBICT972563  and {\em Human Capital and Mobility Program} (Icarus II
project at CINECA, Bologna), and from Japaneese NEDO. JJS's work has been 
supported by the {\it Comunidad Aut\'{o}noma de Madrid}
and the DGICyT through Grants  07T/0024/1998 and  No. PB98-0464.
Computational resources have been
availed from {\em Dep. de F\'{\i}sica de la Materia Condensada} at UAM, 
{\it Abdus Salam International Center for Theoretical Physics} (ICTP) in Trieste,
and JRCAT in Tsukuba, we are deeply thankful to all of them.

\end{multicols} 

\end{document}